\documentclass{notices}
\usepackage{amsmath}
\usepackage{amssymb}
\usepackage{amsthm}
\usepackage{amscd}
\usepackage{bm}
\usepackage{color}
\usepackage{graphicx}
\usepackage{amscd}

\theoremstyle{plain}
\newtheorem{thm}{Theorem}[section]

\theoremstyle{definition}

\theoremstyle{conjecture}

\theoremstyle{problem}

\newcommand{\group}[1]{\mathrm{#1}}
\newcommand{\Hilb}[1]{\mathcal{#1}}
\newcommand{\Sym}[1]{\mathrm{S}_{#1}}
\newcommand{\Hecke}{\mathcal{H}}
\newcommand{\Ho}[1]{\mathrm{H}_{#1}}
\newcommand{\Pair}[1]{\mathcal{M}_{#1}}

\newcommand{\bd}[1]{\mathbf{#1}}  
\renewcommand{\matrix}[1]{\mathbf{#1}}

\newcommand{\End}{\operatorname{End}}

\newcommand{\Fun}{\operatorname{Fun}}
\newcommand{\Par}{\mathsf{Par}}
\newcommand{\type}{\mathsf{type}}

\newcommand{\UU}[1]{\mathrm{U}({#1})}

\newcommand{\WgO}{\operatorname{Wg}^{\mathrm{O}}}
\newcommand{\OO}[1]{\mathrm{O}({#1})}
\newcommand{\WgOO}[1]{\operatorname{Wg}^{\OO{#1}}}
\newcommand{\sgn}{\operatorname{sgn}}

\newcommand{\WgSp}{\operatorname{Wg}^{\mathrm{Sp}}}
\newcommand{\Sp}[1]{\mathrm{Sp}({#1})}
\newcommand{\WgSSp}[1]{\operatorname{Wg}^{\Sp{#1}}}

\newcommand{\trans}[1]{{#1}^{\mathrm{T}}}

\DeclareMathOperator{\Tr}{Tr}

\title{The Weingarten Calculus}

\author{
  Beno\^\i{}t Collins
  \affil{
    Benoit Collins is a professor of mathematics at Kyoto University. His email address is collins@math.kyoto-u.ac.jp
    }
  \and
  Sho Matsumoto
  \affil{
    Sho Matsumoto is an associate professor of mathematics at Kagoshima University. His email address is shom@sci.kagoshima-u.ac.jp
   }
   \and
  Jonathan Novak
  \affil{
    Jonathan Novak is an associate professor at UC San Diego. His email
    address is jinovak@ucsd.edu
   }
}

\date{\empty}

\begin{document}

\maketitle

\section{Introduction}
Every compact topological group supports a unique translation 
invariant probability measure on its 
Borel sets --- the \emph{Haar measure}. 
Haar measure was first constructed for certain families
of compact matrix groups by Hurwitz in the nineteenth century
in order to produce invariants of these groups by averaging
their actions. Hurwitz's construction has been 
reviewed from a modern perspective by Diaconis and Forrester, 
who argue that it should be regarded as the starting point
of modern random matrix theory \cite{MR3612265}.
An axiomatic construction
of Haar measures in the more general context of locally 
compact groups was published by Haar in the 1930s, with 
further important contributions made in work of 
von Neumann, Weil, and Cartan; see \cite{MR2098271}.

Given a measure, one wants to integrate. The 
Bochner integral for continuous functions $F$
on a compact group $\group{G}$ taking 
values in a given Banach space is called the 
\emph{Haar integral}; it is almost always written
simply 

    \begin{equation*}
        \int_{\group{G}} F(g)\, \mathrm{d}g,
    \end{equation*}
    
\noindent
with no explicit notation for the Haar measure.
While integration on groups is a concept of fundamental importance in 
many parts of mathematics, including functional analysis 
and representation theory, probability and ergodic theory,
etc.,
the actual computation of Haar integrals is a problem
which has received curiously little attention.
As far as the authors are aware, it was first considered
by theoretical physicists in the 1970s in the context
of nonabelian gauge theories, where the issue of
evaluating --- or at least approximating --- 
Haar integrals plays a major role. 
In particular, the physics literature on quantum 
chromodynamics, the main theory of strong interactions
in particle physics, is littered with so-called 
``link integrals,'' which are Haar integrals of
the form 

    \begin{equation*}
        \int_{\group{U}(N)} 
        \overline{U_{i(1)j(1)} \dots U_{i(d)j(d)}} 
        U_{i'(1)j'(1)} \dots U_{i'(d)j'(d)}\mathrm{d}U,
    \end{equation*}
    
\noindent
where $\group{U}(N)$ is the compact group of unitary 
matrices $U=[U_{xy}]_{x,y=1}^N.$
Confronted with a paucity of existing mathematical tools for the 
evaluation of such integrals, physicists developed
their own methods, which allowed them to obtain
beautiful, explicit formulas such as

    \begin{equation*}
        \int_{\group{U}(N)}
        \overline{U_{11}U_{22}U_{33}} U_{12}U_{23}U_{31}\mathrm{d}U
        = \frac{2}{N(N^2-1)(N^2-4)},
    \end{equation*}
    
\noindent
an evaluation which holds for all unitary groups of 
rank $N \geq 3.$ Although exceedingly clever, the bag of tricks 
for evaluating Haar integrals assembled by physicists
is ad hoc and piecemeal, lacking the unity and coherence
which are the hallmarks of a mathematical theory.

The missing theory of Haar integrals 
began to take shape in the early 2000s, 
driven by an explosion of interest in
random matrix theory. The basic Hilbert 
spaces of random matrix theory are 
$L^2(\group{H}(N),\text{Gauss})$ and 
$L^2(\group{U}(N),\text{Haar}),$
where $\group{H}(N)$ is the noncompact
abelian group of Hermitian 
matrices $H=[H_{xy}]_{x,y=1}^N$ equipped with a Gaussian measure
of mean $\mu=0$ and variance $\sigma>0$, and $\group{U}(N)$ is 
the compact nonabelian group of
unitary matrices $U=[U_{xy}]_{x,y=1}^N$ equipped with the Haar measure,
just as above. Given a distribution on matrices,
the basic goal of random matrix theory is to understand 
the induced distribution of eigenvalues, which in the 
selfadjoint case form a random point process on the line,
and in the unitary case constitute a random point process on 
the circle. The \emph{moment method} in random matrix theory,
pioneered by Wigner (\cite{MR95527}) in the 1950s,
is an algebraic approach to this problem. The main idea is to
adopt the algebra $\mathcal{S}$ of symmetric polynomials in 
eigenvalues as a basic class of test functions, and integrate
such functions by realizing them as elements of the algebra
$\mathcal{A}$ of polynomials in matrix elements, which can
then (hopefully) be integrated by leveraging the defining features of the matrix
model under consideration. The canonical example is sums 
of powers of eigenvalues, which may equivalently be viewed
as traces of matrix powers; more generally, all coefficients
of the characteristic polynomial are sums of principal matrix
minors.

It is straightforward to see that, in both of the above
$L^2$-spaces, the algebra $\mathcal{A}$ of polynomial 
functions in matrix elements admits the orthogonal decomposition 

    \begin{equation}
    \label{eqn:Orthogonality}
        \mathcal{A} = \bigoplus_{d=0}^\infty \mathcal{A}^{[d]},
    \end{equation}
    
\noindent
where $\mathcal{A}^{[d]}$ is the space of homogeneous degree
$d$ polynomial functions in matrix elements. Thus, modulo the 
algebraic issues inherent in transitioning from $\mathcal{S}$ to $\mathcal{A},$
the moment method boils down to computing scalar products of monomials
of equal degree, so expressions of the form

    \begin{equation*}
        \left\langle \prod_{x=1}^d H_{i(x)j(x)}, \prod_{x=1}^d H_{i'(x)j'(x)} 
        \right\rangle_{L^2(\group{H}(N),\text{Gauss})}
    \end{equation*}

\noindent
and 

    \begin{equation*}
        \left\langle \prod_{x=1}^d U_{i(x)j(x)}, \prod_{x=1}^d U_{i'(x)j'(x)} 
        \right\rangle_{L^2(\group{U}(N),\text{Haar})}.
    \end{equation*}
    
In the Gaussian case, monomial scalar products can be computed
systematically using a combinatorial algorithm
which physicists call the ``Wick formula'' and statisticians 
call the ``Isserlis theorem.'' 
This device leverages independence together with the characteristic feature
of centered normal distributions --- vanishing of all 
cumulants but the second --- to 
compute Gaussian expectations as polynomials in the variance
parameter $\sigma$. The upshot is that
scalar products in $L^2(\group{H}(N),\text{Gauss})$ are closely related 
to the combinatorics of graphs drawn on compact Riemann surfaces,
which play the role of Feynman diagrams for selfadjoint matrix-valued 
field theories. We recommend (\cite{MR1492512})
as an entry point into the fascinating combinatorics of 
Wick calculus.

The case of Haar unitary matrices is a priori more complicated:
the random variables $\{U_{xy} \colon  x,y \in [N]\}$ are 
identically distributed, thanks to the invariance of Haar measure,
but they are also highly correlated, due to the constraint
$U^*U=I.$ Moreover, each individual entry follows a complicated 
law not uniquely determined by its mean and variance. 
Despite these obstacles, it turns out that, 
when packaged correctly, the invariance of Haar measure provides everything needed to
develop an analogue of Wick calculus for Haar unitary matrices. 
Moreover, once the correct general perspective has been found,
one realizes that it applies equally well to any compact group,
and even to compact symmetric spaces and compact 
quantum groups. This compact group analogue of Wick calculus has come to be known as
\emph{Weingarten calculus}, a name chosen by Collins \cite{MR1959915} to 
honor the contributions of Donald Weingarten, a physicist whose
early work in the subject proved to be of foundational importance.

The Weingarten calculus has matured rapidly over the course of
the past decade, and the time now seems right 
to give a pedagogical account of the subject. The authors are
currently preparing a monograph intended to meet this need. 
In this article, we aim to provide an easily 
digestible and hopefully compelling 
preview of our forthcoming work, emphasizing the big picture. 

First and foremost, we wish to impart 
the insight that, like the calculus of Newton and Liebniz,
the core of Weingarten calculus is a fundamental theorem
which converts a computational problem into a symbolic problem:
whereas the usual fundamental theorem of calculus converts
the problem of integrating functions on the line into computing
antiderivatives, the fundamental theorem of Weingarten calculus
converts the problem of integrating functions on groups into 
computing certain matrices associated to tensor invariants.
The fundamental theorem of Weingarten calculus is presented in detail in Section 2.

We then turn to examples illustrating the fundamental theorem
in action. We present two
detailed case studies: integration on the automorphism group 
$\group{S}(N)$ of $N$ distinct points,
and integration on the automorphism group $\group{U}(N)$
of $N$ orthonormal vectors. These are natural examples,
given that the symmetric group and the unitary group are
model examples of a finite and infinite compact group, 
respectively. The $\group{S}(N)$ case, presented in Section 3, is a toy example
chosen to illustrate how Weingarten calculus works in an elementary situation
where the integrals to which it applies can easily be evaluated from 
first principles.
The $\group{U}(N)$ case, discussed in Section 4, is an example of real interest,  
and we give a detailed workup showing how 
Weingarten calculus handles the link integrals of $\group{U}(N)$
lattice gauge theory. 

Section 5 gives a necessarily brief discussion of Weingarten calculus
for the remaining
classical groups, namely the orthogonal group $\group{O}(N)$ and the symplectic
group $\group{Sp}(N),$ both of which receive a detailed treatment in 
(cite our book). Finally, Section 6 extols the universality of 
Weingarten calculus, briefly discussing how it can be transported to
compact symmetric spaces and compact quantum groups, and indicating
applications in quantum information theory.

\section{The Fundamental Theorem}
Given a compact group $\group{G}$, a finite-dimensional
Hilbert space $\Hilb{H}$ with a specified orthonormal
basis $e_1,\dots,e_N$, and a continuous group homomorphism
$U \colon \group{G} \to \group{U}(\Hilb{H})$, let 
$U_{xy} \colon \group{G} \to \mathbb{C}$ be the 
corresponding matrix element functionals,

    \begin{equation*}
        U_{xy}(g) = \langle e_x, U(g) e_y \rangle, \quad 1 \leq x,y \leq N.
    \end{equation*}

\noindent
The \emph{Weingarten integrals} of the unitary representation
$(\Hilb{H},U)$ are the integrals

    \begin{equation*}
        I_{ij} = \int_{\group{G}} \prod_{x=1}^d U_{i(x)j(x)}(g)
        \mathrm{d}g,
    \end{equation*}
    
\noindent
where $d$ ranges over the set $\mathbb{N}$ of
positive integers, and the multi-indices $i,j$ range over
the set $\Fun(d,N)$ of functions from $[d]=\{1,\dots,d\}$
to $[N]=\{1,\dots,N\}$. Clearly, if we can compute all Weingarten
integrals $I_{ij}$, then we can integrate any function 
on $\group{G}$ which is a polynomial in the matrix 
elements $U_{xy}.$ This is
the basic problem of Weingarten calculus: 
compute the Weingarten integrals of a given unitary 
representation of a given compact group.

The fundamental theorem of Weingarten calculus addresses this problem by linearizing it.
The basic observation is that, for each 
$d \in \mathbb{N},$ the $N^{2d}$ integrals $I_{ij}$, $i,j \in \Fun(d,N)$, 
are themselves the matrix elements of a linear operator. 
Indeed, we have

    \begin{equation*}
        I_{ij} = \int_\group{G} U^{\otimes d}_{ij}(g) \mathrm{d}g,
    \end{equation*}
    
\noindent
where

    \begin{equation}
    \label{eqn:TensorBasis}
        e_i = e_{i(1)} \otimes \dots \otimes e_{i(d)}, \quad i \in \Fun(d,N)
    \end{equation}
    
\noindent
is the orthonormal basis of $\Hilb{H}^{\otimes d}$ corresponding to the 
specified orthonormal basis $e_1,\dots, e_N$ in $\Hilb{H}$, and 

    \begin{equation*}
        U_{ij}^{\otimes d}(g) = \langle e_i,U^{\otimes d}(g)e_j \rangle,
        \quad i,j \in \Fun(d,N),
    \end{equation*}
    
\noindent
are the matrix elements of the unitary operator $U^{\otimes d}(g)$ in 
this basis. By continuity, we thus have that

    \begin{equation*}
        I_{ij} = P_{ij}, \quad i,j \in \Fun(d,N),
    \end{equation*}
    
\noindent
where $P_{ij} = \langle e_i,P e_j \rangle$ are the 
matrix elements of the selfadjoint operator 

    \begin{equation*}
        P = \int_\group{G} U^{\otimes d}(g) \mathrm{d}g
    \end{equation*}
    
\noindent
obtained by integrating the unitary operators $U^{\otimes d}(g)$ against
Haar measure.
The basic problem of Weingarten calculus is thus equivalent
to computing the matrix elements of
$P \in \End \Hilb{H}^{\otimes d}$, for all $d \in \mathbb{N}.$

This is where the characteristic feature of Haar measure,
the invariance

    \begin{equation*}
        \int_{\group{G}} F(g_0g) \mathrm{d}g = \int_{\group{G}} F(gg_0) \mathrm{d}g
        = \int_{\group{G}} F(g) \mathrm{d}g, \quad g_0 \in \group{G},
    \end{equation*}
    
\noindent
comes into play: it forces $P^2=P.$ Thus $P$ is a selfadjoint
idempotent, and as such $P$ orthogonally projects $\Hilb{H}^{\otimes d}$
onto its image, which is the space of $\group{G}$-invariant
tensors in $\Hilb{H}^{\otimes d},$
            
        \begin{equation*}
            (\Hilb{H}^{\otimes d})^\group{G} = \{ t \in \Hilb{H}^{\otimes d} \colon
            U^{\otimes d}(g)t = t \text{ for all }g \in \group{G}\}.
        \end{equation*}
 
\noindent  
Thus, we see that the basic problem of Weingarten calculus is in fact 
very closely related to the basic problem of invariant theory, which is 
to determine a basis for the space $\group{G}$-invariant tensors in 
$\Hilb{H}^{\otimes d}$ for all $d \in \mathbb{N}.$

Indeed, suppose we have have access to a basis $a_1,\dots,a_m$ of 
$(\Hilb{H})^{\otimes d}$.
Then, by elementary linear algebra, we have everything 
we need to calculate the matrix

    \begin{equation*}
        \matrix{P} = [I_{ij}]_{i,j \in \Fun(d,N)}
    \end{equation*}

\noindent
of degree $d$ Weingarten integrals.
Let $\matrix{A}$ be the $N^d \times m$ matrix whose 
columns are the coordinates of the basic invariants
in the desired basis, 

    \begin{equation*}
        \matrix{A} = [\langle e_i,a_x \rangle ]_{i \in \Fun(d,N), x \in [m]}.
    \end{equation*}

\noindent
Then we have the matrix factorization

    \begin{equation*}
    \label{eqn:MatrixFactorization}
        \matrix{P}=\matrix{A}(\matrix{A}^*\matrix{A})^{-1}\matrix{A}^*,
    \end{equation*}
    
\noindent
familiar from matrix analysis as the multidimensional
generalization of the undergraduate ``outer product divided by inner product''
formula for orthogonal projection onto a line. The $m \times m$
matrix $\matrix{A}^*\matrix{A}$ is nothing but the Gram matrix 

    \begin{equation*}
    \label{eqn:GramMatrix}
        \matrix{A}^*\matrix{A} = [\langle a_x,a_y \rangle]_{x,y \in [m]}
    \end{equation*}

\noindent
of the basic $\group{G}$-invariants in $\Hilb{H}^{\otimes d}$, 
whose linear independence is equivalent to the invertibility of the Gram matrix.
Let us give the inverse Gram matrix a name: we call 

    \begin{equation*}
    \label{eqn:WeingartenMatrix}
        \matrix{W} = (\matrix{A}^*\matrix{A})^{-1}
    \end{equation*}
    
\noindent
the \emph{Weingarten matrix} of the invariants $a_1,\dots,a_m.$
Extracting matrix elements on either side of the factorization
$\matrix{P}=\matrix{A}\matrix{W}\matrix{A}^*$, we obtain the Fundamental
Theorem of Weingarten Calculus.

	\begin{thm}
	\label{thm:Fundamental}
		For any $d \in \mathbb{N}$ and $i,j \in \Fun(d,N),$ we have
		
		\begin{equation*}
			I_{ij}
			= \sum_{x,y=1}^m \matrix{A}_{ix} \matrix{W}_{xy}
			\matrix{A}^*_{yj}.
		\end{equation*}
			
	\end{thm}
	
Does Theorem \ref{thm:Fundamental} actually solve the 
basic problem of Weingarten calculus? Yes, insofar as the classical fundamental theorem 
of calculus solves the problem of computing definite integrals: 
it reduces a numerical problem to a symbolic problem. In order
to apply the fundamental theorem of calculus to integrate
a given function, one must find its antiderivative, 
and as every student of calculus knows this can be 
a wild ride. In order to use the fundamental theorem of Weingarten
calculus to compute the Weingarten integrals of a given unitary 
representation, one must solve a suped up version of the basic problem of invariant
which involves not only finding basic tensor invariants, but computing 
their Weingarten matrices. Just like the computation
of antiderivatives, this may prove to be a difficult task.

\section{The Symmetric Group}
In this Section, we consider a toy example. Fix 
$N \in \mathbb{N},$ and let $\group{S}(N)$ be 
the symmetric group of rank $N,$ viewed as the 
group of bijections $g \colon [N] \to [N].$
This is a finite group, its topology and resulting Haar measure 
are discrete, and all Haar integrals are finite sums. 
We will solve the basic problem of Weingarten calculus for 
the permutation representation of $\group{S}(N)$ in two ways:
using elementary combinatorial reasoning, and using the fundamental theorem of 
Weingarten calculus. It is both instructive and psychologically reassuring
to work through the two approaches and see that they 
agree.

The permutation representation of $\group{S}(N)$ is the unitary
representation $(\Hilb{H},U)$ in which
$\Hilb{H}$ is an $N$-dimensional Hilbert space with
orthonormal basis $e_1,\dots,e_N,$ and
$U \colon \group{S}(N) \to \group{U}(\Hilb{H})$ 
is defined by 

	\begin{equation*}
		U(g)e_x = e_{g(x)}, \quad x \in [N].
	\end{equation*}
	
\noindent
The corresponding system of matrix elements
$U_{xy} \colon \group{S}(N) \to \mathbb{C}$
is given by

    \begin{equation*}
        U_{xy}(g) = \langle e_x, U(g) e_y \rangle = \delta_{xg(y)},
        \quad x,y \in [N].
    \end{equation*}
    
\noindent
We will evaluate the Weingarten integrals 
of $(\Hilb{H},U),$ 

    \begin{equation*}
        I_{ij} = \int_{\group{S}(N)} 
        \prod_{x=1}^d U_{i(x)j(x)}(g) \mathrm{d}g.
    \end{equation*}

\noindent
Each Weingarten integral $I_{ij}$ is a finite sum with $N!$ terms,
each equal to zero or one:

	\begin{equation*}
	\begin{split}
		I_{ij} 
		&= \frac{1}{N!} \sum_{g \in \group{S}(N)} 
		    \prod_{x=1}^d U_{i(x)j(x)}(g) \\
		&= \frac{1}{N!}\sum_{g \in \group{S}(N)}
		    \prod_{x=1}^d \delta_{g^{-1}i(x),j(x)}
	\end{split}
	\end{equation*}
	
\noindent
Thus, $N! I_{ij}$ simply counts permutations $g \in \group{S}(N)$ 
which solve the equation $g^{-1}i=j.$ 
This is an elementary counting problem,
and a good way to solve it is to think of
the given functions $i,j \in \Fun(d,N)$
``backwards,'' as the ordered lists of their fibers:

	\begin{equation*}
	\begin{split}
		i &= (i^{-1}(1),\dots,i^{-1}(N)) \\
		j &= (j^{-1}(1),\dots,j^{-1}(N)).
	\end{split}
	\end{equation*}
	
\noindent
The fiber fingerprint of the composite function 
$g^{-1}i \in \Fun(d,N)$ is then

	\begin{equation*}
		g^{-1}i = (i^{-1}(g(1)),\dots,i^{-1}(g(N))),
	\end{equation*}
	
\noindent
and so we have $g^{-1}i=j$ if and only if

	\begin{equation*}
		(i^{-1}(g(1)),\dots,i^{-1}(g(N))) = (j^{-1}(1),\dots,j^{-1}(N)).
	\end{equation*}
	
\begin{figure}
        \centering
        \includegraphics[width=8cm]{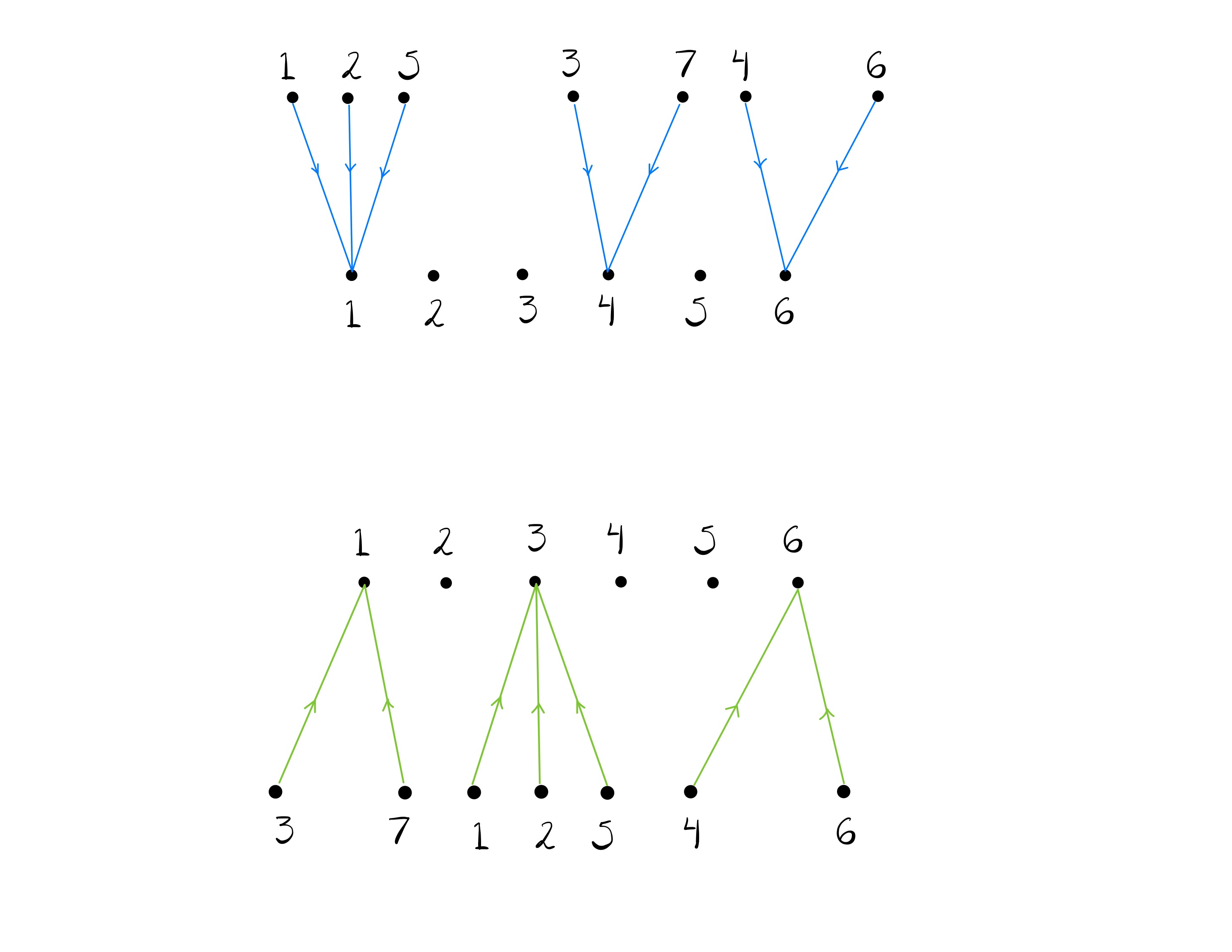}
        \caption{Two functions of the same type}
        \label{fig:type}
    \end{figure}

\noindent
Clearly, such a permutation exists if and only if the fibers of $i$ and $j$ are
the same up to the labels of their base points, which is the case if and only if

	\begin{equation*}
		\type(i) = \type(j),
	\end{equation*}
	
\noindent
where $\type(i)$ is the partition of $[d]$ obtained by forgetting the 
order on the fibers of $i$ and throwing away empty fibers; see Figure
\ref{fig:type}.
When this is the case, the permutations we wish to count number

	\begin{equation}
		\delta_{\mathsf{type}(i)\mathsf{type}(j)}(N - \# \mathsf{type}(i))!
	\end{equation}
	
\noindent
in total, where $\# \pi$ denotes the number of blocks of the set partition $\pi$. 
We conclude that the integral $I_{ij}$ is given by

	\begin{equation}
	\begin{split}
		I_{ij} 
		&=  \delta_{\mathsf{type}(i)\mathsf{type}(j)} \frac{(N - \# \mathsf{type}(i))!}{N!} \\
		&= \frac{\delta_{\mathsf{type}(i)\mathsf{type}(j)}}{N(N-1) \dots (N - \# \mathsf{type}(i)+1)}.
	\end{split}
	\end{equation}

Let us now evaluate $I_{ij}$ using the 
Fundamental Theorem of Weingarten Calculus.
The first step is to solve the basic problem of 
invariant theory for the representation $(\Hilb{H},U).$
This is straightforward. Fix $d \in \mathbb{N},$
let $\mathsf{Par}_N(d)$ denote the set 
of partitions of $[d]$ with at most $N$
blocks, and to each $\mathsf{p} \in \Par_N(d)$ associate the tensor

	\begin{equation*}
		a_{\mathsf{p}} = \sum_{\substack{i \in \Fun(d,N) \\
		\type(i) = \mathsf{p}}} e_{i},
	\end{equation*}
	
\noindent
where $e_i = e_{i(1)} \otimes \dots \otimes e_{i(d)} \in 
\Hilb{H}^{\otimes d}.$
It is apparent that the set 
$\{a_{\mathsf{p}} : \mathsf{p} \in \Par_N(d)\}$ is
a basis of $(\Hilb{H}^{\otimes d})^{\group{S}(N)}$.
Indeed, taking the unit tensor

	\begin{equation*}
		e_i = e_{i(1)} \otimes \dots \otimes e_{i(d)}
	\end{equation*}
	
\noindent
corresponding to a function $i \in \Fun(d,N)$ and 
symmetrizing it using the action of permutations 
on multi-indices produces the tensor

		\begin{equation*}
			a_i = \sum_{g \in \group{S}(N)} 
			e_{gi(1)} \otimes \dots \otimes e_{gi(d)},
		\end{equation*}
	
\noindent
which is clearly $\group{S}(N)$-invariant, and 
moreover it is clear that every $\group{S}(N)$-invariant
tensor in $\Hilb{H}^{\otimes d}$ is a linear combination
of tensors of this form. Furthermore,

	\begin{equation*}
		a_i = a_j \iff \type(i) = \type(j),
	\end{equation*}
	
\noindent
so that the distinct invariants produced
by symmetrization of the initial basis 
in $\Hilb{H}^{\otimes d}$ are 

	\begin{equation*}
		a_{\mathsf{p}} = \sum_{\substack{i \in \Fun(d,N) \\
		\type(i) = \mathsf{p}}} e_i, 
		\quad \mathsf{p} \in \Par_N(d).
	\end{equation*}

\noindent
These tensors are pairwise orthogonal: 
for any $\mathsf{p},\mathsf{q} \in \Par_N(d)$,
we have

		\begin{align*}
			\langle a_{\mathsf{p}},a_{\mathsf{q}} \rangle
			&= \left\langle
				\sum_{i \in \Fun(d,N)} \delta_{\type(i)\mathsf{p}}\vec{e}_i,
				\sum_{j \in \Fun(d,N)} \delta_{\type(j)\mathsf{q}} \vec{e}_j
				\right\rangle \\
				&= \sum_{i \in \Fun(d,N)} \sum_{j \in \Fun(d,N)}
					\delta_{\type(i)\mathsf{p}}\delta_{\type(j)\mathsf{q}}
					\delta_{ij} \\
				&=\delta_{\mathsf{p}\mathsf{q}} \sum_{i \in \Fun(d,N)} 
					\delta_{\type(i)\mathsf{p}} \\
				&= \delta_{\mathsf{p}\mathsf{q}}N(N-1) \dots (N-\#(\mathsf{p})+1).
		\end{align*}
	
\noindent
So, the Gram matrix of the basis $\{a_{\mathsf{p}} \in \Par_N(d)\}$ 
is diagonal, and the corresponding Weingarten $\matrix{W}$ has entries 

    \begin{equation*}
        \matrix{W}_{\mathsf{p}\mathsf{q}} = 
        \frac{\delta_{\mathsf{p}\mathsf{q}}}{N(N-1) 
        \dots (N-\#(\mathsf{p})+1)}.
    \end{equation*}
    
\noindent
We can now apply the fundamental Theorem of Weingarten 
calculus, and doing so we obtain

		\begin{align*}
			I_{ij} &= \int_{\group{S}(N)} 
			U_{i(1)j(1)} \dots U_{i(d)j(d)} \mathrm{d}g \\
			&=\sum_{\mathsf{p},\mathsf{q} \in \Par_N(d)}
			\langle e_i,a_\mathsf{p} \rangle
				\matrix{W}_{\mathsf{p}\mathsf{q}}
			\langle a_\mathsf{q},e_j \rangle \\
			&= \sum_{\mathsf{p},\mathsf{q} \in \Par_N(d)}
				\frac{\delta_{\type(i),\mathsf{p}}
				\delta_{\mathsf{p}\mathsf{q}}
				\delta_{\mathsf{q},\type(j)}}{N(N-1) \dots (N-\#(\mathfrak{p})+1)}
				 \\
			&= \frac{\delta_{\mathsf{type}(i)\mathsf{type}(j)}}
			{N(N-1) \dots (N - \# \mathsf{type}(i)+1)}.
		\end{align*}

\section{The Unitary Group}
In this section we consider a case of real interest: 
integration on the unitary group $\group{U}(N),$
the automorphism group of a system of $N$ orthonormal
vectors $e_1,\dots,e_N$ spanning a Hilbert space $\Hilb{H}$.
The most obvious unitary representation of
this group is the tautological 
representation $(\Hilb{H},U),$ in which $U(g)=g.$
The resulting system of matrix elements 
$U_{xy} \colon \group{G} \to \mathbb{C}$ is then
simply 

    \begin{equation*}
        U_{xy}(g) = \langle e_x,ge_y\rangle, \quad 1 \leq x,y \leq N,
    \end{equation*}
    
\noindent
and it turns out that all corresponding Weingarten integrals

    \begin{equation*}
        I_{ij} = \int_{\group{U}(N)} \prod_{x=1}^d U_{i(x)j(x)}(g)\mathrm{d}g
    \end{equation*}
    
\noindent
vanish. To see this,  
let $\lambda_0$ be an arbitrary complex number
of modulus one, and let $g_0 \in \group{U}(N)$
be the scalar operator with eigenvalue $\lambda_0.$
We then have $U_{xy}(gg_0)=\lambda_0 U_{xy}(g),$ so 
invariance of Haar measure implies
$I_{ij}=\lambda_0^dI_{ij},$ which forces $I_{ij}=0.$

The basic problem of Weingarten calculus becomes much
more interesting when when we replace the tautological
representation with the adjoint representation. 
The carrier space of the adjoint
representation is the algebra $\End \Hilb{H}$ of all linear maps $A \colon 
\Hilb{H} \to \Hilb{H}$ equipped with the 
Hilbert-Schmidt scalar product

    \begin{equation*}
        \langle A,B \rangle = \Tr A^*B,
    \end{equation*}
    
\noindent
and the action $V$ of $\group{U}(N)$ on this Hilbert
space is conjugation, 

    \begin{equation*}
        V(g)A = g^{-1}Ag.
    \end{equation*}
    
\noindent
The orthonormal basis $e_1,\dots,e_N$ in $\Hilb{H}$ 
induces an orthonormal basis in $\End \Hilb{H}$
consisting of the $N^2$ matrix units defined by
    
    \begin{equation*}
        E_{xx'} e_z = e_x \langle e_{x'},e_z \rangle, \quad x,x',z \in [N].
    \end{equation*}
    
\noindent
The matrix units relate the scalar product on 
$\End \Hilb{H}$ to that on $\Hilb{H}$ via

    \begin{equation*}
        \langle E_{xx'},A \rangle = \langle e_x,Ae_{x'} \rangle.
    \end{equation*}
    
\noindent
The matrix elements of the adjoint representation are thus related 
to those of the tautological representation by 

    \begin{align*}
       V_{yy'xx'}(g)&= \langle E_{yy'},V(g)E_{xx'}\rangle \\
        &= \langle ge_y, E_{xx'}g e_{y'} \rangle \\
        &= \sum_z \langle ge_y,E_{xx'}e_z \rangle \langle e_z,ge_{y'} \rangle \\
        &= \sum_z \langle ge_y,e_x \rangle \langle e_{x'},e_z\rangle \langle e_z,ge_{y'} \rangle \\
        &= \overline{U_{xy}(g)} U_{x'y'}(g).
    \end{align*}

\noindent
So, the Weingarten integrals

    \begin{equation*}
        I_{jj'ii'} = \int_{\group{U}(N)} \prod_{x=1}^d V_{j(x)j'(x)i(x)i'(x)}(g) \mathrm{d}g,
    \end{equation*}

\noindent
of the adjoint representation of $\group{U}(N)$ are exactly the link integrals

    \begin{equation*}
        L_{ii'jj'}=\int_{\group{U}(N)} \prod_{x=1}^d \overline{U_{i(x)i'(x)}(g)} U_{j(x)j'(x)}(g) \mathrm{d}g
    \end{equation*}
    
\noindent
of $\group{U}(N)$ lattice gauge theory: we have $I_{jj'ii'}=L_{ii'jj'}.$
        
    \subsection{The Gram matrix}
    In order to calculate Weingarten integrals of the 
    adjoint representaiton of $\group{U}(N),$ we first need to 
    solve the basic problem of invariant theory for this 
    representation. A partial solution to this problem is well-known, and 
    part of a classical circle of ideas, commonly known 
    as Schur-Weyl duality, which relate the representation 
    theory of $\group{U}(N)$ to representations of the symmetric
    groups $\group{S}(d),$ $d \in \mathbb{N}.$
    In particular, it is known that, after identifying 
    $(\End \Hilb{H})^{\otimes d}$ with $\End \Hilb{H}^{\otimes d},$
    the space of $\group{U}(N)$-invariants is spanned by the 
    operators which act by permuting tensor factors,
    
        \begin{equation*}
            A_\pi v_1 \otimes \dots \otimes v_d = 
            v_{\pi(1)} \otimes \dots \otimes v_{\pi(d)},
            \quad \pi \in \group{S}(d).
        \end{equation*}
        
    \noindent
    Moreover, it is not difficult to compute the scalar
    product of any two of these operators: given 
    $\rho,\sigma \in \group{S}(d),$ one finds that 
    
        \begin{equation*}
            \langle A_\rho,A_\sigma \rangle = N^{\#\mathsf{cycles}(\rho^{-1}\sigma)},
        \end{equation*}
        
    \noindent
    where $\#\mathsf{cycles}(\pi)$ is the number of factors 
    in any factorization of $\pi$ into disjoint cyclic permutations,
    so that the Gram matrix of these invariants is the 
    $d! \times d!$ matrix

        \begin{equation*}
            \matrix{A}^*\matrix{A} = \begin{bmatrix}
            N^{\#\mathsf{cycles}(\rho^{-1}\sigma)}
            \end{bmatrix}_{\rho,\sigma \in \group{S}(d)}.
        \end{equation*}
        
    The reason we refer to this as a partial solution 
    to the basic problem of invariant theory for the 
    adjoint representation of $\group{U}(N)$ is that,
    although $\{A_\pi \colon \pi \in \group{S}(d)\}$
    is a spanning set of invariants, it is only a
    basis in the \emph{stable range}, where $1 \leq d \leq N.$
    In the unstable range, $d > N$, the operators $A_\pi$
    are linearly dependent, and their Gram matrix is 
    singular. A satisfactory patch for this issue
    was found relatively recently by Baik and Rains
    (cite "Increasing subsequences and the classical groups"),
    who showed that $\{A_\pi \colon \pi \in \group{S}_N(d)\}$
    is always a basis, where $\group{S}_N(d) \subseteq \group{S}(d)$
    is the set of permutations of $[d]$ with no decreasing subsequence
    of length $N+1.$ Thus, the Gram matrix which needs to 
    be inverted in order to calculate the degree $d$ Weingarten
    integrals of the adjoint representation is actually 
    
        \begin{equation*}
            \matrix{A}^*\matrix{A} = \begin{bmatrix}
            N^{\#\mathsf{cycles}(\rho^{-1}\sigma)}
            \end{bmatrix}_{\rho,\sigma \in \group{S}_N(d)}.
        \end{equation*}
        
    In the unstable range, the Gram matrix $\matrix{A}^*\matrix{A}$
    must be computed numerically, but in the stable range we can 
    view $N$ as a parameter, so that the Weingarten matrix 
    $\matrix{A}=(\matrix{A}^*\matrix{A})^{-1}$ is a $d! \times d!$
    matrix whose entries are rational functions of $N.$ 
    To get a handle on what these functions might be, 
    it turns out to be a good idea to reinterpret 
    the Gram matrix from the viewpoint of geometric group theory.
    More precisely, let us identify $\group{S}(d)$ with its (right) Cayley
    graph as generated by the conjugacy class of transpositions;
    then, the geodesic distance between permutations 
    $\rho,\sigma \in \group{S}(d)$ is given by 
    $|\rho^{-1}\sigma|,$ where 
    
        \begin{equation*}
            |\pi| = d - \#\mathsf{cycles}(\pi)
        \end{equation*}
        
    \noindent
    is the word norm corresponding to the generating set of 
    transpositions. Let $q$ be a complex parameter, and consider
    the $d! \times d!$ matrix
    
        \begin{equation*}
             \Gamma = \begin{bmatrix}
                 {} & \vdots & {} \\
                \dots & q^{|\rho^{-1}\sigma|} & \dots \\
                {} & \vdots & {}
                \end{bmatrix}_{\rho,\sigma \in \group{S}(d)},
        \end{equation*}
        
    \noindent
    the $q$-distance matrix of the symmetric group $\group{S}(d).$
    The $q$-distance matrix $\Gamma$ is a natural analytic 
    continuation of the Gram matrix $\matrix{A}^*\matrix{A}$ ---
    to recover the latter from the former, simply multiply by
    $q^{-d}$ and then set $q=\frac{1}{N}.$ 
    
    Thus, the problem we face is that of understanding the 
    $q$-distance matrix of the symmetric group sufficiently
    well that we can invert it. This may be addressed via harmonic analysis on 
    $\group{S}(d).$ The basic observation is that $\Gamma$
    is the matrix of the group algebra element
    
        \begin{equation*}
            \gamma = \sum_{\pi \in \group{S}(d)} q^{|\pi|} \pi
        \end{equation*}
        
    \noindent
    acting in the right regular representation of 
    $\mathbb{C}\group{S}(d).$ Moreover, $\gamma$ is a central element 
    in $\group{S}(d)$: in fact, we have
    
        \begin{equation*}
            \gamma = \sum_{r=0}^{d-1} q^r L_r,
        \end{equation*}
        
    \noindent
    where $L_r$ is the sum of all 
    points on the sphere of radius $r$ centered at the 
    identity permutation $\iota \in \group{S}(d),$ or 
    equivalently the sum of all permutations on the 
    $r$th level of the Cayley graph. Clearly, every 
    such sphere/level is a disjoint union of conjugacy 
    classes. The plan is thus to take the Fourier transform
    of $\gamma$, i.e. its image under the algebra 
    isomorphism
    
        \begin{equation}
            \mathcal{F} \colon \mathbb{C}\group{S}(d) 
            \longrightarrow \bigoplus_{\lambda \vdash d}
            \End \mathrm{V}^\lambda,
        \end{equation}
        
    \noindent
    where $(\mathrm{V}^\lambda,R^\lambda)$ is the 
    irreducible representation of $\group{S}(d)$
    indexed by a given Young diagram $\lambda$ with
    $d$ cells, and 
    
        \begin{equation*}
            \mathcal{F}(a) = \bigoplus_{\lambda \vdash d} R^\lambda(a),\quad 
            a \in \mathbb{C}\group{S}(d).
        \end{equation*}
        
    \noindent
    Since $\gamma \in \mathbb{C}\group{S}(d)$ is central,
    Schur's Lemma guarantees that $\mathcal{F}(\gamma)$
    will be a direct sum of scalar operators, which can 
    then easily be inverted. In particular, the computation
    reduces to calculating the Fourier transforms
    of the levels $L_r$ of the Cayley graph.
        
    The computation of the Fourier transform of $L_r$ rests
    on a pair of remarkable discoveries in algebraic combinatorics made by the 
    Lithuanian physicist Algimantas Adolfas Jucys 
    (not to be confused with his father, the Lithuanian physicist
    Adolfas Jucys). The first of Jucys' discoveries 
    is a unique factorization theorem 
    for permutations. Let us call a factorization 
    
        \begin{equation*}
            \pi = (i_1\ j_1) \dots (i_r\ j_r)
        \end{equation*}
        
    \noindent
    of a permutation $\pi \in \group{S}(d)$ into 
    transpositions $(i\ j),$ where $1 \leq i<j \leq d$,
    a \emph{strictly monotone factorization} if 
    $j_1 < \dots < j_r.$
    
        \begin{thm}
            Every permutation $\pi \in \group{S}(d)$ admits
            a unique monotone factorization, and the number
            of factors in this factorization is $|\pi|.$
        \end{thm}
    
        \begin{figure}
            \centering
            \includegraphics[width=7cm]{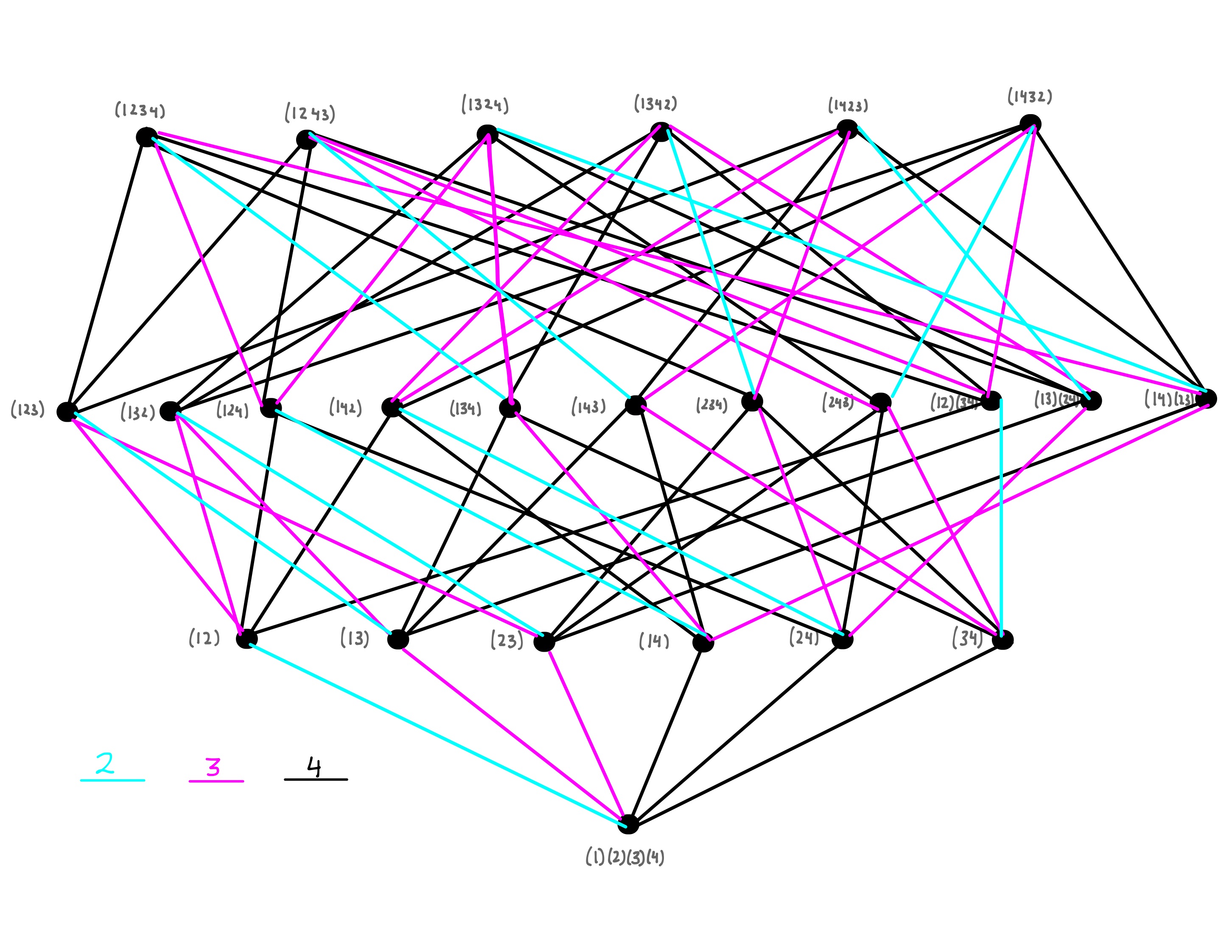}
            \caption{Biane-Stanley labeling of $\group{S}(4)$.}
            \label{fig:BSlabeling}
        \end{figure}
    
\begin{figure}
            \centering
            \includegraphics[width=8cm]{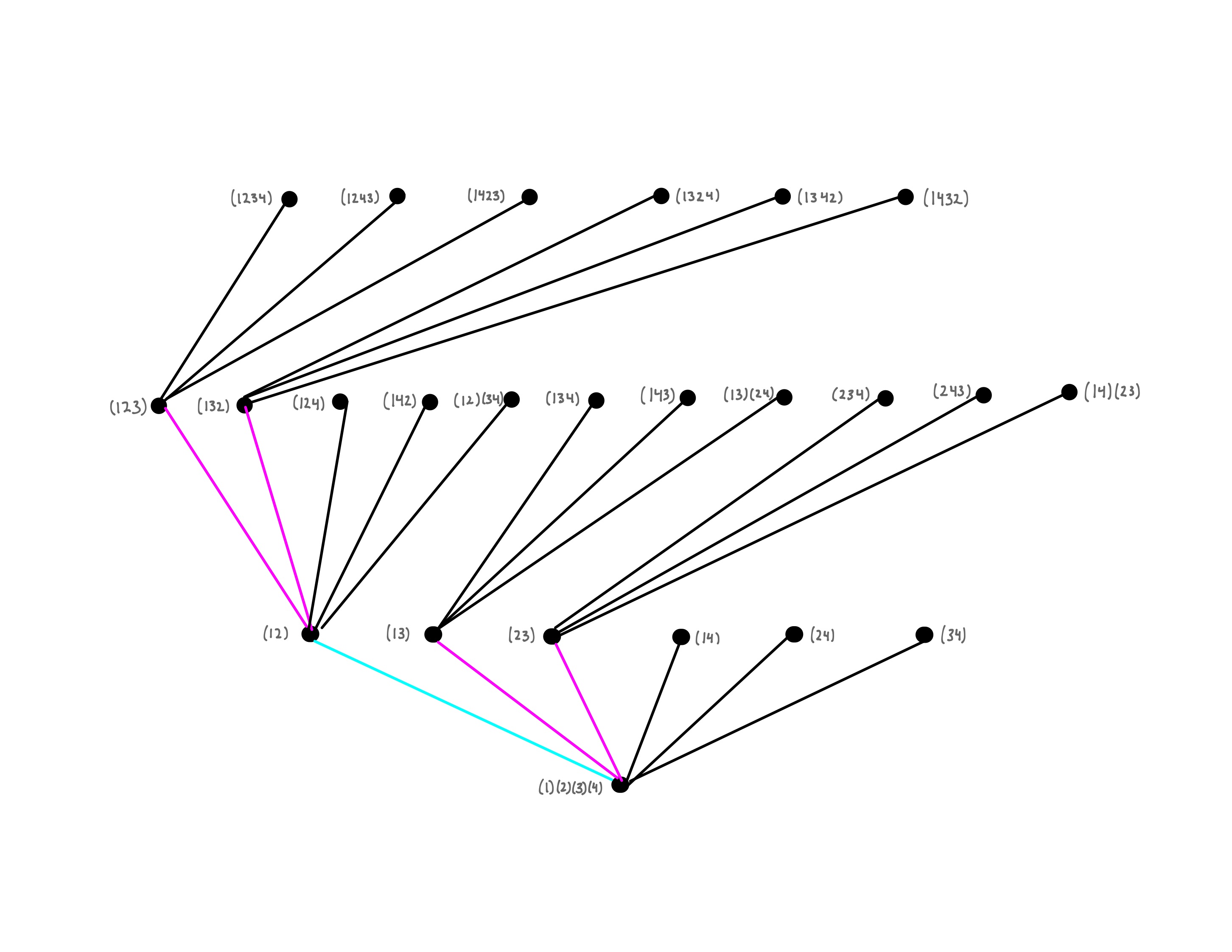}
            \caption{The Jucys tree of $\group{S}(4)$.}
            \label{fig:JucysTree}
        \end{figure}

    This result may be visualized as follows.
    Let us mark each edge of the Cayley graph of 
    $\group{S}(d)$ corresponding to the transposition 
    $(i\ j)$ with $j,$ the larger of the two symbols 
    it interchanges. We call this the \emph{Biane-Stanley
    labeling} of the symmetric group, since a version of 
    it was considered first by Stanley and later by 
    Biane in connection with the combinatorics of 
    noncrossing partitions. Figure \ref{fig:BSlabeling}
    depicts the Biane-Stanley labeling of $\group{S}(4).$
    Call a walk on $\group{S}(d)$
    a \emph{strictly monotone walk} if the labels of the 
    edges it traverse form a strictly increasing 
    sequence. Jucys' result says that if we trace out
    all strictly monotone walks on $\group{S}(d)$ issuing from 
    the identity permutation $\iota,$ we get a presentation of the 
    symmetric group as a starlike tree. Figure \ref{fig:JucysTree}
    depicts the Jucys tree of $\group{S}(4)$.
    
    Jucys' result gives us an a new combinatorial description of the 
    sphere $L_r$: it is the set of all permutations admitting
    a strictly monotone factorization of length $r$, i.e. the
    set of all points at distance $r$ from $\iota$ on the 
    Jucys tree. This in turn gives us a new 
    algebraic description of $L_r$: it may be written as 
    
        \begin{equation*}
            L_r = e_r(J_1,\dots,J_d),
        \end{equation*}
        
    \noindent
    where 
    
        \begin{equation*}
            e_r(x_1,\dots,x_d) = \sum_{\substack{j \in \Fun(r,d) \\
            i \text{ strictly increasing }}} x_{j(1)} \dots x_{j(d)}
        \end{equation*}
        
    \noindent
    is the elementary symmetric polynomial of degree $r,$ and 
    $J_1,\dots,J_d \in \mathbb{C}\group{S}(d)$ are the transposition sums
    
        \begin{equation*}
            J_j = \sum_{i < j} (i\ j), \quad 1 \leq j \leq d.    
        \end{equation*}
        
    \noindent
    These sums are nowadays known as the \emph{Jucys-Murphy} elements
    of $\group{S}(d).$ Although they are clearly non-central, it is 
    not difficult to see that they commute with one another; in fact,
    they generate a maximal abelian subalgebra of $\mathbb{C}\group{S}(d)$
    known as the \emph{Gelfand-Tsetlin subalgebra}, whose role
    in the representation theory of $\group{S}(d)$
    is analogous to the role of maximal tori in Lie theory
    \cite{MR1443185}.
    
    This brings us to Jucys' second discovery. It is a classical 
    result of Newton that the elementary symmetric polynomials
    are algebraically independent and generate the ring of 
    symmetric polynomials. Thus, $f(J_1,\dots,J_d)$ lies in the 
    center of $\group{S}(d)$ for any symmetric polynomial $f$,
    hence $f(J_1,\dots,J_d)$ acts as 
    a scalar operator in any irreducible representation $\mathrm{V}^\lambda.$
    What is its eigenvalue? This question was answered by 
    Jucys in terms of the so-called ``contents'' of Young diagrams:
    if $\Box \in \lambda$ is a cell of the diagram $\lambda,$ its content 
    $c(\Box)$ is simply its column index minus its row index.
    
        \begin{thm}
            For any symmetric polynomial $f$ and any 
            Young diagram $\lambda \vdash d$, we have
            
                \begin{equation*}
                    R^\lambda (f(J_1,\dots,J_d)) = 
                    \omega^\lambda(f)I_{\mathrm{V}^\lambda},
                \end{equation*}
                
            \noindent
            where 
            
                \begin{equation*}
                    \omega^\lambda(f) = f( c(\Box) : \Box \in \lambda)
                \end{equation*}
            
            \noindent
            is the evaluation of $f$ on the multiset of contents of
            $\lambda$ and $I_{\mathrm{V}^\lambda}$ is the 
            identity operator in $\End \mathrm{V}^\lambda.$
        \end{thm}
        
    The above results allow us to compute the Fourier transform
    of $\gamma$: by Jucys' first theorem, we have the factorization,
    
        \begin{equation*}
            \gamma = \sum_{r=0}^d q^r e_r(J_1,\dots,J_d) = 
            \prod_{k=1}^d (\iota + qJ_k),
        \end{equation*}
        
    \noindent
    and hence by Jucys' second theorem we have
    
        \begin{equation*}
            \mathcal{F}(\gamma) = \bigoplus_{\lambda \vdash d} \omega^\lambda(\gamma)
            I_{\mathrm{V}^\lambda},
        \end{equation*}
        
    \noindent
    where 
    
        \begin{equation*}
            \omega^\lambda(\gamma) = \prod_{\Box \in \lambda} (1 + qc(\Box)).
        \end{equation*}
        
    \noindent
    This leads immediately to the conclusion that
    $\gamma \in \mathbb{C}\group{S}(d)$ is 
    invertible for $|q| < \frac{1}{d-1}$, and
    that the Fourier transform of its inverse 
    is 
    
        \begin{equation*}
            \mathcal{F}(\gamma^{-1}) = \bigoplus_{\lambda \vdash d}
            \omega^\lambda(\gamma^{-1}) I_{\mathrm{V}^\lambda},
        \end{equation*}
        
    \noindent
    where the eigenvalue of $\gamma^{-1}$ acting in 
    $\mathrm{V}^\lambda$ is 
    
        \begin{align*}
            \omega^\lambda(\gamma^{-1}) 
            &=\prod_{\Box \in \lambda} (1 + qc(\Box))^{-1} \\
            &= \sum_{r=0}^\infty (-q)^r h_r(c(\Box) \colon \Box \in \lambda),
        \end{align*}
        
    \noindent
    where 
    
        \begin{equation*}
            h_r(x_1,\dots,x_d) = \sum_{\substack{j \in \Fun(r,d) \\
            i \text{ weakly increasing}}} x_{j(1)} \dots 
            x_{j(d)}
        \end{equation*}
        
    \noindent
    is the complete homogeneous symmetric polynomial of
    degree $r.$
    
    \subsection{The Weingarten Matrix}
    The preceding Fourier analysis of the 
    $q$-distance matrix of $\group{S}(d)$ 
    allows us to make a number of powerful
    statements about the Weingarten matrix 
    $\matrix{W}$ of the $\group{U}(N)$-invariants
    $A_\pi \in \End \Hilb{H}^{\otimes d}$, 
    in the stable range $1 \leq d \leq N.$
    
    The first such statement says that we 
    can calculate the entries of the $d! \times d!$
    matrix $\matrix{W}$ explicitly provided we 
    have access to the character table of $\group{S}(d)$.
    
        \begin{thm}
            For any $\rho,\sigma \in \group{S}(d),$ we have
            that 
            
                \begin{equation*}
                    \matrix{W}_{\rho\sigma} = 
                    \sum_{\lambda \vdash d} 
                    \frac{\chi^\lambda(\rho^{-1}\sigma)}{\prod_{\Box \in \lambda}
                    (N+c(\Box))} \frac{\dim \mathrm{V}^\lambda}{d!},
                \end{equation*}
                
            \noindent
            where $\chi^\lambda$ is the character of $\mathrm{V}^\lambda.$
        \end{thm} 
        
    Note that, since $\chi^\lambda(\rho^{-1}\sigma)$ depends only 
    on the cycle type $\alpha$ of the product $\rho^{-1}\sigma$,
    i.e. the Young diagram whose row lengths encode the lengths 
    of the disjoint cycles of this permutation, the matrix 
    entry $\matrix{W}_{\rho\sigma}$ itself depends only 
    on $\alpha.$ We may thus define 
    
        \begin{equation*}
            \mathrm{Wg}^{\group{U}(N)}(\alpha) := \matrix{W}_{\rho\sigma},
        \end{equation*}
        
    \noindent
    this being a function on Young diagrams known as the
    \emph{Weingarten function} of the unitary group $\group{U}(N)$.
    One also writes $\mathrm{Wg}^{\group{U}(N)}$ when it is convenient
    to view the Weingarten function as a central function on 
    permutations.

    Combining Theorem 4.3
    with the Fundamental Theorem of Weingarten Calculus, we thus 
    obtain the following summation formula for the Weingarten
    integrals of adjoint representation of $\group{U}(N),$ which
    are exactly the link integrals of $\group{U}(N)$ gauge theory.
    
        \begin{thm}
            For any $1 \leq d \leq N$ and any 
            $i,j \in \Fun(d,N)$, we have 
            
                \begin{equation*}
                    I_{ij} = \sum_{\rho,\sigma \in \group{S}(d)} 
                    \delta_{i,i'\rho} \delta_{j,j'\sigma} 
                    \matrix{W}_{\rho\sigma}.
                \end{equation*}
        \end{thm}
        
    To the best of our knowledge, this summation formula first 
    appeared in a 1980 physics paper of Samuel \cite{MR597583};
    it was independently rediscovered by Collins in \cite{MR1959915}.
    The fact that the formula is confined to the stable range 
    $1 \leq d \leq N$ turns out to be a minor issue, and this 
    restriction can be easily lifted (\cite{MR2217291}).
    
    A more serious limitation on the utility of Theorem 4.4 is
    the fact that the characters of $\group{S}(d)$ are not at all
    simple objects; in fact, it is a known theorem of complexity
    theory that the irreducible characters of the symmetric groups
    are computationally intractable. Luckily, for many purposes, in both mathematical physics
    and random matrix theory, it is sufficient to have an asymptotic
    estimate for $I_{ij}$ giving its approximate value as $N \to \infty.$
    It turns out that the Fourier analysis of the $q$-distance matrix 
    discussed above gives a complete $N \to \infty$ asymptotic 
    expansion for the entries of $\matrix{W}$.
    
        \begin{thm}
            In the stable range $1 \leq d \leq N,$ we have 
            
                \begin{equation*}
                    \matrix{W}_{\rho\sigma} = 
                    \frac{(-1)^{|\rho^{-1}\sigma|}}{N^{d+|\rho^{-1}\sigma|}}
                    \sum_{k=0}^\infty \frac{\vec{W}_k(\rho,\sigma)}{N^{2k}},
                \end{equation*}
                
            \noindent
            where $\vec{W}_k(\rho,\sigma)$ is the number of weakly monotone
            walks on $\group{S}(d)$ from $\rho$ to $\sigma$ of length
            $|\rho^{-1}\sigma|+2k.$
        \end{thm}
        
    A weakly monotone walk on the Cayley graph of $\group{S}(d)$ is 
    similar to the strictly monotone walks discussed above,
    the difference being that labels of the edges traversed
    are only required to form a weakly increasing sequence.
    Unlike strictly monotone walks, there exist arbitrarily 
    long weakly monotone walks between any two permutations
    $\rho$ and $\sigma$, though these must satisfy a parity 
    constraint depending on whether $\rho^{-1}\sigma$ is an 
    even or odd permutation; this is why the series in 
    Theorem 4.5 is a power series in $N^{-2}.$
    Theorem 4.5 gives a precise combinatorial interpretation of 
    the famous $1/N$ expansion in $\group{U}(N)$ lattice gauge theory, cf
    \cite{MR2567222}.
    The observation that monotone walks on symmetric groups play the 
    role of Feynman diagrams for Haar integrals on $\group{U}(N)$ was
    first made in \cite{MR2730867}, and further developed
    in \cite{MR3010693}. In particular, 
    the number of weakly monotone geodesics between any 
    pair of permutations may be computed in closed form, 
    giving a very useful first order approximation to the entries of 
    $\matrix{W}.$
        
        \begin{thm}
            For any $\rho,\sigma \in \group{S}(d)$, we have
            
                \begin{equation*}
                    \vec{W}_0(\rho,\sigma) = \prod_{i=1}^{\ell(\alpha)}
                    \frac{1}{\alpha_i}{2\alpha_i \choose \alpha_i},
                \end{equation*}
                
            \noindent
            where $\alpha \vdash d$ is the cycle type of
            $\rho^{-1}\sigma.$
        \end{thm}
        
    Yet another ramification of the realization that monotone walks on 
    $\group{S}(d)$ are the Feynman diagrams for Haar integration on 
    $\group{U}(N)$ is a family of identities that play the role of 
    Schwinger-Dyson ``loop'' equations, and recursively determine
    the Weingarten function. The loop equations for $\mathrm{Wg}^{\group{U}(N)}$
    were first obtained by Samuel \cite{MR597583}, and later rediscovered in 
    \cite{MR3680193}, who used them to obtain estimates
    in the unstable range $d>N.$

\section{Orthogonal and symplectic groups}

In this section, we extend the Weingarten calculus for unitary groups in the previous section to 
orthogonal and symplectic groups.
The theory was first considered in \cite{MR2217291},
and further developed with the use of harmonic analysis of symmetry groups in \cite{MR2567222,
MR3077830}.
Since the Weingarten calculus for 
$\group{O}(N)$ and $\Sp{N}$ is parallel to $\group{U}(N)$, 
we focus on stating the results. 

\subsection{Pairings and hyper-octahedral groups}

We realize the (real) orthogonal group $\OO{N}$ as the compact matrix group consisting of all $N\times N$ real orthogonal matrices $g$, that is $ g \trans{g}= I_N$.
We are interested in the expectation of monomials 
$r_{i(1) j(1)} r_{i(2) j(2)} \dots r_{i(k) j(k)}$
in matrix elements $r_{xy}=\langle \bd{e}_{x}, g \bd{e}_{y} \rangle$
if $g$ is distributed with respect to the Haar probability $\mathrm{d} g$ on $\OO{N}$.

Since two random orthogonal matrices $g$ and $-g$ are distributed in the same law, 
the integral 
$\int_{\OO{N}}  r_{i_1 j_1} r_{i_2 j_2} \dots r_{i_k j_k} \, \mathrm{d}g
=\int_{\OO{N}}  (-r_{i_1 j_1}) (-r_{i_2 j_2}) \dots (-r_{i_k j_k}) \, \mathrm{d} g$ 
vanishes if $k$ is odd,
so we consider only even-degree moments.

To do that, we introduce the notion of pairings and hyper-octahedral groups.
Let $\Pair{2d}$  be the set of all \emph{pairings} of $\{1,2,\dots, 2d\}$, that is,
set partitions of $\{1,2,\dots,2d\}$ whose blocks are size two.
Each pairing $\sigma$ can be expressed  in the form 
$\sigma=\{ \{\sigma(1), \sigma(2)\}, \{\sigma(3), \sigma(4)\},\dots,\{\sigma(2d-1), \sigma(2d)\}\}$,
where $\sigma(1), \sigma(2), \dots, \sigma(2d)$ is a permutation of $1,2,\dots,2n$.
We often write it in the condition 
\begin{gather}
\sigma(2r-1)<\sigma(2r) \quad (1 \le r \le d),  \label{eq:pair_ordered} \\
1=\sigma(1)<\sigma(3)< \dots <\sigma(2d-1), \notag 
\end{gather}
and identity it with a permutation expressed in  the same symbol $\sigma$ in $\Sym{2d}$. 
Namely, we regard $\Pair{2d}$ as a subset of $\Sym{2d}$.
For example, a pairing $\{\{1,5\}, \{2,8\},\{3,4\}, \{6,7\}\}$ is identified with the permutation
$\left(\begin{smallmatrix}
1 & 2 & 3 & 4 & 5 & 6 & 7 & 8 \\
1 & 5 & 2 & 8 & 3 & 4 & 6 & 7  
\end{smallmatrix} \right)$ in $\Sym{8}$.

Let $\Ho{d}$ be the subgroup of $\Sym{2d}$ generated by elements 
$(2i-1,2i)$ with $1\le i \le d$ and $(2i-1,2j-1)(2i,2j)$ with
$1 \le i<j \le d$, where $(p,q)$ stands for the transposition between $p$ and $q$.
We call it the \emph{hyper-octahedral group} of degree $d$. 
The set $\Pair{2d}$, which is regarded as a subset of $\Sym{2d}$, forms a complete set of representatives of left cosets $\sigma \Ho{d}$ in $\Sym{2d}$.

Furthermore, in order to distinguish double cosets $\Ho{d} \sigma \Ho{d}$, we consider an undirected multigraph $\mathbf{\Gamma}(\sigma)$ for each $\sigma \in \Sym{2d}$ as follows.
The vertex set of $\mathbf{\Gamma}(\sigma)$ is $\{1,2,\dots,2d\}$, and the edge set consists of $\{\{2i-1,2i\} \ | \ 1 \le i \le d \}$ and 
$\{\{\sigma(2i-1), \sigma(2i)\} \ | \ 1 \le i \le d\}$.
Each vertex lies on exactly two edges. 
Then connected components of $\mathbf{\Gamma}(\sigma)$ are cycles of even lengths $2\mu_1, 2\mu_2, \dots, 2\mu_l$, where we arrange them with
$\mu_1 \ge \mu_2 \ge \dots \ge \mu_l \ge 1$. 
We call the (integer) partition $\mu=(\mu_1,\mu_2,\dots,\mu_l)$ of $d$ the \emph{coset-type} of $\sigma$. 
For example, for a permutation $\sigma= \left(\begin{smallmatrix}
1 & 2 & 3 & 4 & 5 & 6 & 7 & 8 \\
1 & 5 & 2 & 8 & 4 & 3 & 6 & 7  
\end{smallmatrix} \right)$, one connected component of $\mathbf{\Gamma}(\sigma)$ has six vertices $1,5,6,7,8,2$ and another component has two vertices $3,4$; so 
its coset-type is $\mu=(3,1)$.
It is known that two permutations $\sigma, \tau$ in $\Sym{2d}$ have the same coset-type if and only if they belong to the same double coset of $\Ho{d}$ in $\Sym{2d}$, i.e., $\Ho{d} \sigma \Ho{d}= \Ho{d} \tau \Ho{d}$.
The length $\kappa(\sigma)$ of the coset-type of $\sigma \in \Sym{2d}$
is important. 
Equivalently, it is the number of connected components in the graph $\mathbf{\Gamma}(\sigma)$.

\medskip

\subsection{Weingarten formula for orthogonal groups}

Now we give Weingarten formula for the orthogonal group $\OO{N}$.
Let $\bd{i}=(i(1),i(2), \dots,i(2d))$ and $\bd{j}=(j(1),j(2),\dots,j(2d))$
be sequences of length $2d$ whose entries picked up from $\{1,2,\dots,N\}$.
Then we have the formula
\begin{multline} \label{eq:WgFormula-Ortho}
\int_{\OO{N}} r_{i_1 j_1} r_{i_2 j_2} \dots r_{i_{2d}, j_{2d}} \, \mathrm{d} g  \\
=\sum_{\sigma \in \Pair{2d}} \sum_{\tau \in \Pair{2n}} 
\Delta_{\sigma}(\bd{i})\Delta_{\tau}(\bd{j}) \WgOO{N}(\sigma^{-1}\tau),
\end{multline}
where $\Delta_{\sigma}(\bd{i})$ is, by definition,  equal to $1$ if $i(a)=i(b)$ for every pair $\{a,b\}$ in $\sigma$; 
to zero otherwise.
We here skip a detailed definition of  $\WgOO{N}$, 
which can be obtained by the same argument as in the case of unitary groups,
but we look at a few examples first.
For each permutation $\sigma$, the value $\WgOO{N}(\sigma)$ depends on only its coset-type.
We denote by $\sigma_\mu$ a specific permutation with coset-type $\mu$. 
Then we may see that 
\begin{align}
\WgOO{N}(\sigma_{1})=& \frac{1}{N}, \\
\WgOO{N}(\sigma_{1,1})=& \frac{N+1}{N(N-1)(N+2)}, \label{eq:WgO-11} \\
\WgOO{N}(\sigma_{2})=& \frac{-1}{N(N-1)(N+2)}. \label{eq:WgO-2}
\end{align}

Let us see an application for formula \eqref{eq:WgFormula-Ortho}. 
Consider two sequences $\bd{i}=(1,1,2,2)$ and $\bd{j}=(2,3,2,3)$.
Then $\Delta_{\sigma}(\bd{i})=1$ only if $\sigma=\{\{1,2\},\{3,4\}\}$;  $\Delta_{\tau}(\bd{j})=1$ only if $\tau=\{\{1,3\},\{2,4\}\}$.
When we regard these $\sigma,\tau$ as permutations, the coset-type of $\sigma^{-1} \tau$ is the same with that of $\sigma_2$.
Thus, we obtain the integral value
\begin{multline*}
 \int_{\OO{N}} r_{12} r_{13} r_{22} r_{23} \, \mathrm{d}g \\
 =  \WgOO{N}(\sigma_{2})  = \frac{-1}{N(N-1)(N+2)}.
\end{multline*}

The discussion of orthogonal Weingarten functions can be almost parallel to that of unitary cases, 
but in a slightly more complicated form.
For example, the counterpart of 
the $1/N$ expansion of the unitary Weingarten function is as follows:
for any $1 \le d \le \frac{N+1}{2}$ and  any $\alpha \vdash d$,
we have
    \begin{equation*}
        \WgOO{N}(\sigma_\alpha)= \frac{(-1)^{d-\ell(\alpha)}}{N^{2d-\ell(\alpha)}}
        \sum_{k=0}^\infty (-1)^k\frac{\vec{W}'_k(\alpha)}{N^k},
    \end{equation*}
where $\vec{W}'_k(\alpha)$ is a non-negative integer 
enumerating certain analogues of monotone walks on $\Pair{2d}$.

\subsection{Weingarten formula for symplectic groups}

Let $J=J_N$ be the $2N \times 2N$ skew symmetric matrix given by
\begin{equation} \label{eq:matrix-J}
J_N= 
\begin{pmatrix} O & I_N \\ - I_N & O  \end{pmatrix}.
\end{equation}
The (unitary) symplectic group $\Sp{N}$ is realized as 
$\Sp{N}=\{g \in \UU{2N} \ | \ \trans{S} J S=J\}$.
This preserves the skew symmetric bilinear form on $\mathbb{C}^{2N}$ given by
$\langle \bd{v},\bd{w} \rangle_J = \trans{\bd{v}} J \bd{w}$.
If the collection $\{\bd{e}_1,\dots, \bd{e}_{2N}\}$ is the standard basis of $\mathbb{C}^{2N}$, then
it is immediate to see that 
\[
\langle \bd{e}_i, \bd{e}_j \rangle_J= 
\begin{cases} 1 & \text{if $j=i+N$}, \\
-1 & \text{if $i=j+N$}, \\
0 & \text{otherwise}.
\end{cases}
\]

The Weingarten formula for $\Sp{N}$ is quite similar to $\OO{N}$ but we need to treat signatures carefully.
Consider the integral 
$\int_{\Sp{N}}  s_{i(1) j(1)} s_{i(2) j(2)} \cdots s_{i(k), j(k)} \, \mathrm{d} g$
of matrix elements,  where $\mathrm{d} g$ is the Haar probability on $\Sp{N}$.
As for the orthogonal groups, this integral vanishes if $k$ is odd.
Here we use matrix elements $s_{xy}$ of $g$ rather than the value of the skew form
$\langle \bd{e}_x, g \bd{e}_y \rangle_J$.

For each pairing $\sigma \in \Pair{2d}$ and a sequence $\bd{i}=(i(1),i(2),\dots,i(2d))$
of length $2d$ picked up from $\{1,2,\dots,2N\}$,
we define
\begin{equation*}
\Delta_{\sigma}' (\bd{i})= \prod_{r=1}^d \langle \bd{e}_{i(\sigma(2r-1))}, \bd{e}_{i(\sigma(2r))} \rangle_J.
\end{equation*}
This Delta-symbol takes the value of $1$, $-1$, or $0$.
Here we must watch the assumption \eqref{eq:pair_ordered};
otherwise, the sign of this may  be  accidentally changed.

Now we provide Weingarten formula for symplectic groups.
For 
two sequences $\bd{i}=(i(1),i(2), \dots,i(2d))$ and $\bd{j}=(j(1),j(2),\dots,j(2d))$
picked up from $\{1,2,\dots,2N\}$, 
we have 
\begin{multline} \label{eq:WgFormula-Sp}
\int_{\Sp{N}} s_{i(1) j(1)} s_{i(2) j(2)} \dots s_{i(2d) j(2d)} \, \mathrm{d} g \\
=\sum_{\sigma \in \Pair{2d}} \sum_{\tau \in \Pair{2d}} 
\Delta_{\sigma}'(\bd{i})\Delta_{\tau}'(\bd{j}) \WgSSp{N}(\sigma^{-1}\tau).
\end{multline}

Let us see an example for symplectic Weingarten formula \eqref{eq:WgFormula-Sp}.
Consider the integral 
$\int_{\Sp{N}} s_{1,1} s_{2,N+2} s_{N+1,2} s_{N+2,N+1} \, \mathrm{d}g$, so we apply  \eqref{eq:WgFormula-Sp} with
$\bd{i}=(1,2,N+1,N+2)$ and $\bd{j}=(1,N+2, 2,N+1)$.
Then only parings $\sigma=\{\{1,3\},\{2,4\}\}$ and $\tau=\{\{1,4\},\{2,3\}\}$ contribute to the sum in  \eqref{eq:WgFormula-Sp},
and we have 
$\Delta'_\sigma(\bd{i})=\langle \bd{e}_{1}, \bd{e}_{N+1} \rangle_J \langle \bd{e}_{2}, \bd{e}_{N+2} \rangle_J = +1$ and 
$\Delta'_\tau(\bd{j})=\langle \bd{e}_{1}, \bd{e}_{N+1} \rangle_J \langle \bd{e}_{N+2}, \bd{e}_{2} \rangle_J =-1$. 
Moreover, the permutation $\sigma^{-1} \tau$ is 
$\left(\begin{smallmatrix} 1 & 2 & 3 & 4 \\ 1 & 3 & 2 & 4 \end{smallmatrix} \right)^{-1}
\left(\begin{smallmatrix} 1 & 2 & 3 & 4 \\ 1 & 4 & 2 & 3 \end{smallmatrix} \right)
= \left(\begin{smallmatrix} 1 & 2 & 3 & 4 \\ 1 & 4 & 3 & 2 \end{smallmatrix} \right)$,
which is of sign $-1$ and of coset-type $(2)$.
In the present text, we do not give the definition of 
the symplectic Weingarten function, but
such an observation show that 
the integral is equal to
\begin{multline*}
\int_{\Sp{N}} s_{1,1} s_{2,N+2} s_{N+1,2} s_{N+2,N+1} \, \mathrm{d}g \\
=\WgSSp{N}(\sigma_{2})=\frac{1}{4N(N-1)(2N+1)}.
\end{multline*}

\subsection{Circular ensembles}

In random matrix theory, not only classical compact groups $\UU{N}, \OO{N}, \Sp{N}$ but also circular ensembles are well studied.
The three main examples are circular orthogonal/unitary/symplectic ensembles (COE/CUE/CSE).
In this subsection, we will follow the symbols of Random Matrix Theory and regard random matrices as matrix-valued random maps, and write integrals $\int \cdots \mathrm{d} g$
in the form of expectation values $\mathbb{E}[ \cdots ]$.

The CUE matrix is nothing but the Haar-distribited unitary matrix, the Weingartn calculus for which is already given in the previous section. 
Let $U$ and $\tilde{U}$ be two CUE matrices of dimension $N$ and $2N$, respectively.
Then the COE matrix $V=(v_{ij})_{i,j=1}^N$ and CSE matrix $\tilde{H}=(\tilde{h}_{ij})_{i,j=1}^{2N}$ are determined by
$V=U \trans{U}$ and $\tilde{H}= \tilde{U} J \trans{\tilde{U}} \trans{J}$, with the matrix $J$ defined in \eqref{eq:matrix-J}, respectively.
However, for a technical reason, we consider a modified CSE matrix $H= \tilde{U} J \trans{\tilde{U}}$ rather than $\tilde{H} = H \trans{J}$.

The Weingarten formulas for them are given as follows.
We denote by $\mathbb{E}$ the corresponding expectation for each random matrix.
For two sequences 
$\bd{i}=(i(1), i(2), \dots, i(2m))$ and $\bd{j}=(j(1),j(2),\dots, j(2n))$,
whose entries are picked up from $\{1,2,\dots,N\}$,
we have the formula for the COE
\begin{multline} \label{eq:Wg_COE}
\mathbb{E}\big[v_{i(1) i(2)} v_{i(3) i(4)} \dots v_{i(2m-1) i(2m)} \\ \overline{v_{j(1) j(2)} v_{j(3) j(4)} \dots v_{j(2n-1) j(2n)} } \big] \\
= \delta_{mn} \sum_{\sigma \in \Sym{2n}} \delta_{\sigma}(\bd{i},\bd{j}) \WgO(\sigma;N+1).
\end{multline}
Similarly, for two sequences 
$\bd{i}$ and 
$\bd{j}$ from $\{1,2,\dots,2N\}$,
we have the formula for the CSE
\begin{multline*}
\mathbb{E}\big[h_{i(1) i(2)} h_{i(3) i(4)} \dots h_{i(2m-1) i(2m)} \\ 
\overline{h_{j(1) j(2)} h_{j(3) j(4)} \dots h_{j(2n-1) j(2n)} } \big] \\
= \delta_{mn}\sum_{\sigma \in \Sym{2n}} \delta_{\sigma}(\bd{i},\bd{j}) \WgSp(\sigma;N-\tfrac{1}{2}).
\end{multline*}
Here $\delta_{\sigma}(\bd{i},\bd{j})$ is, by definition,  equal to $1$ if $i(\sigma(r))=j(r)$ for all $r \ge 1$; 
to zero otherwise.
Moreover, $\WgO(\sigma;z)$ and $\WgSp(\sigma;z)$ are the rational function in $z$, obtained $N$ by a complex number $z$ 
for $\WgOO{N}(\sigma)$ and $\WgSSp{N}(\sigma)$, respectively.

Surprisingly, when we think of COE and CSE, 
we do not need any new Weingarten function, 
but a different parameter of the orthogonal/symplectic Weingarten functions suffice.

The COE and CSE are deeply related to compact symmetric spaces 
$\UU{N}/\OO{N}$ and $\UU{2N}/\Sp{N}$, respectively. 
For other kinds of compact symmetric spaces, with corresponding various random matrices, 
similar rich Weingarten formulas are known.

Historically, the formula \eqref{eq:Wg_COE} first appeared in \cite{MR1411614} without proof.
Mathematical treatment for COE and other compact symmetric spaces were done in \cite{MR2967964, MR3077830}.

\section{Conclusion and Outlook}
In this article, we have only scratched the surface of Weingarten calculus,
both in terms of theory and applications. 

On the theoretical side, the results
we have presented for integration on $\group{U}(N)$, and only touched on for 
$\group{O}(N)$ and $\group{Sp}(N)$, can be rendered in much more detail and
admit many powerful generalizations which we have not discussed here. Moreover,
the entire apparatus can be developed in the context of compact symmetric spaces
and compact quantum groups, where the results are just as rich and varied as for
classical compact topological groups. We 
touched on Weingarten calculus for symmetric spaces when discussing
circular ensembles of random matrices above, and here we will briefly 
indicate the situation for compact quantum groups.
Roughly speaking, compact quantum groups are noncommutative $C^*$-algebras 
obtained from the $C^*$-algebras of classical compact 
topological groups by suppressing commutativity.
They enjoy the same key properties as the function algebras of 
classical compact groups, namely they satisfy a Peter-Weyl theorem, 
a Tannaka-Krein duality, they admit a finite left and right invariant
Haar measure, and all their irreducible representations are of finite dimension.
The theory was created by Woronowicz, who laid these foundations in a series of
landmark papers. A version of the Weingarten calculus for the 
computation of Haar integrals on compact quantum group was derived in 
\cite{MR2341011}, as an extension of the works of \cite{MR1959915}, and has 
since found many applications in functional analysis and operator algebras.
Our forthcoming monograph gives the first pedagogical account of this new
theory.

Concerning applications of the Weingarten calculus, there are many.
Historically, one of the first applications of Weingarten
calculus is a systematic approach to asymptotic
freeness of random matrices, a phenomenon discovered by 
Voiculescu in the context of free probability theory, see e.g.
\cite{MR1217253}.
Roughly speaking, free probability theory is a noncommutative
probability theory in which the notion of independence is 
based on the free product of algebras, as opposed to the 
tensor product, which gives classical independence. This
notion arises naturally in the study of certain von Neumann
algebras, but Voiculescu discovered that large, classically 
independent random matrices in fact approximate free random
variables. 
We refer to \cite{MR3585560} for references. 
This fact is enormously useful in random matrix
theory, as it allows the machinery of free probability theory
to be harnessed in order to study the asymptotic spectral 
behavior of families of large random matrices. Initially,
the connection between random matrices and free probability 
was only applicable to global observables of the spectrum,
such as expectations of traces of powers as discussed earlier.
It turns out that, when the machinery of Weingarten calculus is 
brought into the picture, it becomes possible to amplify this
connection to \emph{strong asymptotic freeness}, which enables
the use of free probability methods to handle non-global observables,
such as the operator norm of random matrices. It turns out that
this boost is precisely what is needed to bring the tools of 
random matrix theory and free probability to bear on theoretical
problems in quantum information theory (\cite{MR3432743}).

\bibliography{biblio.bib}

\end{document}